 \documentclass[reqno, 12pt]{article}
\usepackage{a4wide,amssymb}
\usepackage{amsthm,amsfonts,amsmath}
\usepackage{graphicx
,
}
\usepackage{eucal}
\usepackage{enumerate}

 \newcommand{\nn}{\nonumber}
\newcommand{\R}{{\mathbb R}}

\renewcommand{\d}{\delta}
           
\newcommand{\var}{\varepsilon}
\newcommand{\lam}{\lambda}

\newcommand{\pa}{\partial}

\newcommand{\varf}{\varphi}

\let\D=\Delta   \let\G=\Gamma

\let\e=\varepsilon

\begin{document}


\begin{center} 
{\bf \Large{Propagation of Chaos and Effective Equations in Kinetic Theory: a Brief Survey}}

\vspace{1.5cm}
{\large M. Pulvirenti$^{1}$ and S. Simonella$^{2}$}

\vspace{0.5cm}
{$1.$\scshape {\small \ Dipartimento di Matematica, Universit\`a di Roma La Sapienza\\ 
Piazzale Aldo Moro 5, 00185 Roma -- Italy \\ \smallskip
$2.$\ Zentrum Mathematik, TU M\"{u}nchen \\ Boltzmannstrasse 3, 85748 Garching -- Germany}}
\end{center}

\vspace{1.5cm}
\noindent
{\bf Abstract.} 
We review some historical highlights leading to the modern perspective on the concept of chaos from the point of view of the kinetic theory. We focus in particular on the role played by the propagation of chaos in the mathematical derivation of effective equations.


%
%
%
%
%
%

\section*{1. The paradigm of the kinetic theory}

\noindent
Propagation of chaos is a central topic in kinetic theory and certainly
exhibits interesting features from the point of view of probability theory and mathematical physics.

This work is dedicated to our friend and colleague Lucio Russo, who gave and is giving important 
contributions to these fields and to the history of science. The purpose is to review some important steps 
in the mathematical understanding of kinetic equations and of the notion of chaos.

We do not pretend to be exhaustive and limit ourselves to a selection of arguments
which played a key role from a modern outlook. We also comment on some perhaps less known historical 
aspects underlining the long and difficult path of the scientific progress.

Many interesting systems in physics and applied sciences are
constituted by a large number of identical components so that they are
difficult to analyze mathematically. On the other
hand, quite often, we are not interested in a detailed description of the
system but rather in his collective behaviour. Therefore it is
necessary to look for all procedures leading to simplified models,
retaining the interesting features of the original system, cutting
away redundant informations. This is exactly the methodology of
statistical mechanics and kinetic theory. Here we want 
 to outline the limiting procedure leading from the microscopic description of a large particle
system (based on the fundamental laws like the Newton or the Schr\"odinger
equation) to the more practical picture dictated by the kinetic theory. 

Although recently the methodology of the kinetic theory has been applied to a large variety of complex systems 
(constituted by a huge number of individuals), we will discuss here only models arising in physics, and more precisely in classical mechanics.
The starting point is
a system of $N$ identical particles in the space $\Bbb R^3$. A microscopic state of the system is a sequence $z_1,\cdots, z_N$ where $z_i=(x_i,v_i)$ denotes position and velocity of the $i$th particle.
The particles interact via the (smooth) two-body interaction $\varphi : \Bbb R^3 \to \Bbb R$ and the equations of motion are

\begin{equation}
\begin{cases}
\displaystyle \dot x_i=v_i  \\
\displaystyle  \dot v_i =- \sum_{j: j\neq i }Ê\nabla \varphi (x_i-x_j)
\end{cases}\;.\label{new}
\end{equation}
Particles have unit mass and $\varphi$ depends on the distance $|x_i-x_j|$ so that the force of the particle $j$ acting on 
the particle $i$ (that is $-\nabla \varphi (x_i-x_j)$) is directed along $ x_i-x_j $.

We are interested in a situation where $N$ is very large (for instance a cubic centimeter or a rarefied gas contains approximately $10^{19}$ molecules). The knowledge of the microscopic states becomes useless and we turn to a statistical description. We introduce a probability measure  $W^N (Z_N) dZ_N$ (absolutely continuous with respect to the Lebesgue measure), defined on the phase space of the system $\Bbb R^{3N}\times \Bbb R^{3N}$
where $$Z_N=(z_1, \cdots, z_N)=(x_i,v_i, \cdots, x_N, v_N)\;.$$
$W^N$ assigns the same statistical weight to two different vectors $Z_N$ and $Z'_N$ differing only for the order of particles, i.e.\;identifying 
the same physical configuration. Physically relevant measures are symmetric with respect to permutations of the sequence
$z_1, \cdots, z_N$.

The time evolved measure is defined  by
\begin{equation}\label {meas}
W^N(Z_N,t)=W^N( \Phi ^{-t }(Z_N) ).
\end{equation}
Here $\Phi ^{t }(Z_N)$ denotes the  the dynamical flow constructed by solving the equations of motion, namely  $\Phi ^{t }(Z_N)$ solves  \eqref {new} with initial datum $Z_N$.

 We can establish a partial differential equation, called Liouville equation, describing the evolution of the measure \eqref{meas}. However also this equation is not tractable from a practical point of view.
 To have an efficient reduced description one can focus on the time  evolution for the probability distribution of a {\em given} particle (say particle $1$),  
 all the particles being identical.

To this end we define the $j$-particle marginals  
\begin{equation}\label{marg}
f^N_j(Z_j,t)=\int  dz_{j+1}Ê\cdots dz_{N} W^N (Z_j, z_{j+1},Ê\cdots, z_{N}, t), \qquad j=1, \cdots, N
\end{equation}
and we look for an equation describing the evolution of $f^N_1$.
Roughly, we establish an evolution equation of the form
\begin{equation}\label{hie11}
\pa_t f_1^N  =-v\cdot \nabla f^N_1 + Q\;.
\end{equation}
The first term in the right hand side denotes the contribution to the evolution of $f^N_1$ due to the free transport of particles, while the term $Q$ 
should describe the interaction of particle $1$ with the rest of the system. 

We are led to face a big difficulty.  Since the interaction is binary, $Q$ will depend on $f^N_2$, namely the two-particle marginal. 
In other words, \eqref{hie11} is still useless
because to know $f^N_1$ we need to know  $f^N_2$ and, to know $f^N_2$, we need to know  $f^N_3$... 
We handle a hierarchy of equations, called BBKGY hierarchy \cite{Bog62} (from the names of the physicists Bogolyubov, Born, Green, Kirkwood and Yvon).

Here enters the property called {\em propagation of chaos}, that is:
\begin{equation}
\label{PC1}
f^N_2(x_1,v_1, x_2, v_2, t) = f^N_1(x_1,v_1, t) f^N_1( x_2, v_2, t). 
\end{equation}
Accepting \eqref{PC1}, $Q$ becomes a bilinear operator of $f^N_1$ and \eqref{hie11} is a closed equation.
We have thus replaced a huge ordinary differential system with a single PDE. The price we pay is that \eqref{hie11} is nonlinear.

Strictly speaking, \eqref{PC1} is certainly false, since it expresses the statistical independence of particle $1$ and particle $2$
which, even if assumed at time zero, cannot hold at later times. Indeed the dynamics creates correlations.
Nevertheless, one can hope to recover this property in some {\em asymptotic} situation described by a suitable scaling limit.
This is what happens  in three different physical contexts:
the mean-field, the low-density and the weak-coupling limits,  yielding three different kinetic equations, namely the Vlasov, the Boltzmann and the 
Landau equation respectively.

\section*{2. Mean-Field limit and Vlasov equation}

The simplest example in which the methods of the kinetic theory apply is the mean-field limit.
Let us suppose that the particle system we are considering interacts through a very small (possibly long-range) potential $O(\frac 1N)$,
where the number of particles $N$ is going to diverge. The equation of motion becomes:
\begin{equation}
\begin{cases}
\displaystyle \dot x_i=v_i  \\
\displaystyle  \dot v_i =-  \frac 1N \sum_{j: j\neq i } \nabla \varphi (x_i-x_j).
\end{cases}\;.\label{new1}
\end{equation}
Consider also an initial distribution $W^N$ fully factorized, i.e. $W^N=f_0^{\otimes N}$.
In this situation the dynamics creates correlations at each positive time. However, given two particles, say $1$ and $2$,
the dynamics of particle $1$ is influenced by the presence of particle $2$ by a factor  $O(\frac 1 N)$. The same happens for particle $2$
as regards the influence of particle $1$. Therefore we expect that the correlations are negligible and in the limit $N \to \infty$
\begin{equation}
\label{fact}
f^N_2 \approx  (f_1^N ) ^{\otimes 2}.
\end{equation}
We shall see that, in our context, \eqref{hie11} becomes
\begin{equation}\label{hie1V}
(\pa_t+v_1\cdot \nabla_{x_1}) f_1^N (t) =\frac {(N-1)}N  \int dx_2 \int dv_2 Ê\nabla \varphi (x_1-x_{2}) \cdot \nabla_{v_1} f_{2}^N(x_1,v_1,x_2,v_2,t),
\end{equation}
so that, using \eqref {fact} and taking formally the limit $N \to \infty $, we arrive to the following equation for the one-particle distribution $f$:

\begin{equation}\label{vlasov}
(\pa_t+v \cdot \nabla_{x}) f (x,v, t) = \int dy \int dw f (y,w,t) Ê\nabla \varphi (x-y) \cdot \nabla_{v} f (x,v,t)\;.
\end{equation}
Equation \eqref{vlasov} is called the {\em Vlasov equation} (by the name of the physicist who introduced it) and describes a large system of weakly interacting particles. 

The rigorous analysis of the mean-field limit and the Vlasov equation are a well understood subject in case of smooth potentials (see for instance Dobrushin \cite {Dob79}).
The interesting case of the Coulomb interaction is still a challenging open problem.

\section*{ 3. The Boltzmann equation}

Much more subtle are the limiting physical situations leading to the Boltzmann equation.

Ludwig Boltzmann established an
evolution equation to describe the behaviour of a rarefied gas (1872), starting
from the mathematical model of elastic  balls and using mechanical and statistical
considerations. The importance of this equation is two-fold. From one side it
provides (as well as the hydrodynamical equations) a reduced description of the
microscopic world. On the other, it is also an important tool
for the applications, especially for dilute fluids when the
hydrodynamical equations fail to hold.

According to the general paradigm of the kinetic theory, the starting point of the Boltzmann analysis is to renounce to study the 
 gas in terms of the detailed motion of the molecules of the full system. It is preferable to investigate a function
$f(x,v)$ which is the probability density of a given particle, where $x$ and $v$ denote its position and velocity. 

Following the original approach proposed by Boltzmann,
$f(x,v) dx dv$ is rather to be interpreted as the fraction of molecules falling in the cell
of the phase space of size $dx$ $dv$ around $x,v$. 
The two concepts are not
exactly the same but they are asymptotically equivalent (when the number of
particles diverges) if a {\em law of large numbers} holds. 

More precisely, Boltzmann considers
 the occupation numbers of the cells of a grid  in the phase space,
when the side of the cell is macroscopically small but sufficiently large to contain a huge number of particles.
From an historical point of view the Boltzmann analysis is really remarkable. At that time, probability theory was not 
well developed mathematically and the possibility of describing the macroscopic world 
in terms of atoms and molecules was still doubtful.

Boltzmann considered a gas as microscopically described by a system of elastic (hard) balls,
colliding according to the laws of classical mechanics.
 
The {\em Boltzmann equation} for the one-particle distribution function reads:
\begin{equation}\label{beq}
(\pa_t+v\cdot \nabla_x)f=Q(f,f)
\end{equation}
where $Q$, the collision operator, is defined by 
\begin{equation}\label{bcoll}
Q(f,f)(x,v) = \int_{\R^3} dv_1 \int_{S_+^2} dn  \quad (v-v_1)\cdot n \quad
[f (x,v') f(x,v'_1)-f (x,v)f(x,v_1)]\;,
\end{equation}
\begin{equation*} 
\label{scatt}
v'=v-n[n\cdot(v-v_1)] 
\end{equation*}
\begin{equation}\label{urto}
v_1'=v_1+n[n\cdot(v-v_1)]
\end{equation}
and $n$  is a unitary vector varying in $S_+^2=\{n\ |\ n\cdot(v-v_1)\geq 0\}$.

Note that $ v',v_1'$ are the outgoing velocities for a collision of
two elastic balls with incoming velocities $v$ and $v_1$ and centers $x$ and
$x+\e n$ respectively, being $\e$ the diameter of the spheres. The collision takes
place if $n\cdot(v-v_1)> 0$. Formulas \eqref{urto} are consequence of the conservation of energy and momentum.
Note that $\e$ does not enter \eqref{beq} as a parameter.

As a fundamental feature of \eqref{beq}, one has the formal conservation (in time) of the
following five quantities:
\begin{equation}\label{mass}
\int  dx  \int dv f(x,v,t) v^{\alpha}
\end{equation}
with $\alpha=0,1,2,$ expressing respectively conservation of probability, momentum and
energy.
From now on we shall set $\int=\int_{\R^3}$ for notational
simplicity.

Moreover Boltzmann introduced the (kinetic) entropy defined by 
\begin{equation}\label{entr}
H(f)=\int dx\int dv f\log f (x,v)
\end{equation}
and proved the famous $H$ theorem asserting the decrease of $H(f(t))$ along the
solutions of \eqref{beq}.

Finally, in case of bounded domains or homogeneous solutions ($f=f(v,t)$
is independent of $x$), the distribution defined for
some $\beta >0$, $\rho >0$, and $u\in \R^3$, by
\begin{equation}\label{max}
M(v)=\frac {\rho}{(\frac {2\pi} {\beta})^{3/2}}e^{-\frac {\beta}2 {|v-u|^2}},
\end{equation}
called Maxwellian distribution, is stationary for the evolution given by \eqref{beq}.
In addition $M$ minimizes $H$ among all distributions with given total mass
$\rho$, mean velocity $u$ and mean energy. The parameter $\beta$ is interpreted
as the inverse temperature.

In conclusion Boltzmann was able to introduce an evolutionary equation with the
remarkable property of expressing mass, momentum and energy conservations but
also trend to thermal equilibrium. In other words, he tried to
conciliate the Newton laws with the second principle of thermodynamics. 

The $H$ theorem is in contrast with the  laws of mechanics, which are time reversible. This fact
caused  skepticism among the scientific community and the work of Boltzmann was attacked by several scientists.
We address the reader to the beautiful monograph by C. Cercignani \cite {C98}, which is
a marvelous compromise between history and high level scientific divulgation, to have a faithful idea 
of the debate at the time.

\bigskip

To derive formally \eqref{beq},  let us consider a system of $N$ identical hard spheres of diameter $\e$ and
unitary mass, interacting by  means
of the collision law \eqref{urto}.  We denote by $\e$ the diameter of the particles in view 
of the fact that $\e$ is very small compared with typical macroscopic lengths.



The  phase space $\Gamma_N$ of the system is the subset of    $ (\mathbb{R}^6)^N$ fulfilling the hard-core condition, namely
$$
  |x_i-x_j| \geq \e    \quad \text{for }i \neq j .
$$

The dynamical flow is  defined as the free flow, i.e.
$$Z_N(t)= (x_1+v_1t ,v_1,\cdots ,x_N+v_Nt,v_N)$$ up to the first impact time, namely when
$|x_i-x_j|=\e$ for some $i \neq j$. 
Then an instantaneous collision takes place in accord to the law \eqref{urto} and the free 
flow goes on up to the next collision instant.

We denote by
$Z_N\to \Phi ^t (Z_N) $ the dynamical flow constructed in this way. The well-posedness of the hard-sphere dynamics is not obvious, due to the occurrence of multiple collisions or to the a priori possibility that collision times accumulate at a finite limiting time. However such pathologies  cannot occur outside a set of initial conditions  $Z_N$ of vanishing measure. Therefore the flow $Z_N\to \Phi ^t (Z_N) $ can be defined almost everywhere with respect to the Lebesgue measure, and this is enough for our purposes. 

Given a probability measure with density $W^\var_0$ on $ \Gamma_N$,  thanks to the invariance of the Lebesgue measure under the above evolution, we define 
the time evolved measure as the measure with density
 \begin{equation}\label{meas1}
W ^\e(Z_N,t) = W^\e_0  (\Phi^{-t}Z_N )\;.
 \end{equation}
 
We recall that we consider probability distributions $W^\e_0 $ which are initially (and hence at any positive time) symmetric in the
exchange of the particles. The probability density of the first $j$ particles is given by the $j$-particle marginal
\begin{equation}\label{marg}
f^\e_j(Z_j,t)=\int  dz_{j+1}Ê\cdots dz_{N} W^\e (Z_j, z_{j+1},Ê\cdots, z_{N}, t), \qquad j=1, \cdots, N\;.
\end{equation}

\medskip
{\em Notational remark.} Up to now the two parameters $N$ an $\e$ have been introduced independently,
and the definitions $\G_N, W^\e, f_j^\e \cdots$ should exhibit a double dependence.
However in a moment we shall fix a precise dependence $\e=\e(N)$ so that the notation is unambiguous.
\medskip

Cercignani \cite{Ce72} derived a hierarchy
of equations for the marginals and the first of such equations, for the one-particle
distribution, $f^\e = f^\e_1$, is  
\begin{equation}\label {beq2}
(\pa_t+v\cdot \nabla_x)f^\e=\hbox {Coll},
\end{equation}
where `Coll' denotes the variation due to the collisions. It takes the form

\begin{equation}\label{bcoll2}
\hbox {Coll}=
(N-1)\e^2 \int dv_2 \int_{S^2} dn \, f^\e_2 (x,v,x+n\e,v_2) (v_2-v)\cdot n.
\end{equation}

Let us argue on the physical significance of \eqref{bcoll2} and \eqref {beq2}. In absence of collisions the probability density of a given particle
would be conserved, that is
$$
\frac d {dt} f^\e( x+vt,v,t)=0.
$$ 
The presence of collisions and the total conservation of the probability imply that
$$
\frac d {dt} f^\e( x+vt,v,t)=\text {flux},
$$
where the `flux' is computed  on the boundary of the spherical surface of the ball
of radius $\e$ around $x$. More precisely, the probability flux  due to a collision with a given particle,
say particle $2$, having velocity $v_2$, is given by
$$
 \e^2 \int_{S^2} f^\e_2 (x,v, x+n\e,v_2) V\cdot n
 $$
 where $V=v_2-v$ is the relative velocity and $-n$ is the inward normal to the considered surface.
 Integrating with respect to $dv_2$ and summing over all the possible choices of particles, we arrive to
 \eqref{beq2} and \eqref{bcoll2}.

 This is, basically, Boltzmann's original argument, except for an important 
 conceptual gap.
 The object of investigation is not the probability density but rather the quantity
 $$
 f^\e(x,v,t) \approx \frac {N_{\D}(t) }{|\D| N}
 $$
 where $\D$ is a ``small''  cell on the phase space around the point $(x,v)$, $|\D|$ its volume and
 $N_{\D}(t) $ is the occupation number of the cell $\D$ (number of particles falling in $\D$)
 at time $t$.
 Clearly, $\D$ must be small compared with the macroscopic lengths, e.g.\;the size of the box in which the gas is confined,
 but large with respect to the typical microscopic lengths, for instance $\e$ (molecular diameter).
 In view of a {\em limit} $N \to \infty$,
 it is possible to conceive a law of large numbers allowing to identify the empirical sample with the a priori probability.

%

Next we shall tackle the problem of getting a closed equation.
Apparently, we are in a situation analogous to the one discussed for the Vlasov equation, but there is a deep difference.
Indeed for the hard-sphere system one can write a hierarchy of equations which plays the role of the BBGKY hierarchy 
for smooth potentials, but the interaction among the particles is strong and the mean-field argument used to invoke the 
propagation of chaos fails.
Boltzmann's most important assumption enters here, namely that two given particles should be (almost) {\em uncorrelated} if the 
gas is {\em rarefied} enough. This leads to the propagation of chaos
\begin{equation}\label{caos}
f^\e_2 (x,v,x_2,v_2) \approx f^\e(x,v) f^\e(x_2,v_2)\;,
\end{equation}
which is however much more delicate in the present context. In fact if two particles collide, correlations are created. 
Even assuming \eqref{caos} at some time, if particle 1 collides with particle 2
such a property cannot be satisfied at any time {\em after} the collision.

Before discussing the propagation of chaos further, we notice that, in practical situations,
for a rarefied gas, the combination $N\e^3
\approx 10^{-8} cm ^3$ (total volume occupied by the particles) is very small, while $N\e^2=O(1)$. 
This implies that the collision operator given by \eqref{bcoll2} is $O(1)$.
Therefore, since we are dealing with a huge number of particles, we are 
tempted to perform the limit
$N\to \infty$ and $\e \to 0$ in such a way that
$\e^2=O(N^{-1})$. As a consequence,  
the probability that two tagged particles collide (which is of the order of the
surface of a ball, that is $O(\e^2)$) is negligible. Instead, the probability that 
a given particle performs a collision with {\em any} of the remaining $N-1$
particles ($O(N\e^2)=O(1)$) is not negligible. On the other hand, condition \eqref {caos}
is referring to two preselected particles (say $1$ and $2$) and
it is not unreasonable to conceive that it holds in the limiting situation in
which we are working.  

Nevertheless, we {\em cannot} insert \eqref {caos} in \eqref {bcoll2} because the integral operator
refers to times both before and after the collision. Let us assume \eqref {caos} only
when the pair of velocities $v,v_2$ are incoming ($(v-v_2 ) \cdot n >0$).
If the two particles are initially uncorrelated, it is unlikely that they have collided before a given time $t$,
 so that we assume  their statistical independence.
 
 This is a standard argument in textbooks of kinetic theory but some extra comment is needed.
 If particle $1$ and $2$  have not collided directly before a given time $t$,  this does not imply that they are uncorrelated. For instance
 it may exist a chain of collisions involving a group $ i_1,  i_2  \cdots  $ of particles
 $$
 1 \to i_1 \to i_2 \to \dots \to 2\;,
 $$
 correlating particles $1$ and $2$. The occurrence of this event must be excluded
 by a rigorous mathematical analysis.


%

Coming back to \eqref {bcoll2}, for the outgoing pair velocities $v,v_2$ ($(v_2-v)\cdot n>0$) 
we make use of the continuity property
\begin{equation}\label{cont}
f^\e_2 (x,v,x+n\e, v_2) =f^\e_2 (x,v',x+n\e,v_2') 
\end{equation}
where the pair $v',v_2'$ is  pre-collisional. On the two-particle distribution expressed in terms of 
pre-collisional variables we apply condition \eqref {caos}, obtaining
\begin{equation*}
\hbox{Coll}=(N-1)\e^2 \int dv_2 \int_{S_+^2} dn  \,(v-v_2)\cdot n
\end{equation*}\label{coll3}
\begin{equation}
\times [f^\e (x,v') f^\e (x-n\e,v'_2)-f^\e(x,v)f^\e(x+n\e,v_2)]
\end{equation}
after a change $n\to -n $ in the positive part of $\hbox{`Coll'}$, using the notation $S_+^2$
for the hemisphere  $S_+^2=\{n\ |\ n\cdot(v-v_2)\geq 0\}$. 

 Finally, in the limit $N\to \infty$, $\e\to 0$, $N\e^2=\lam^{-1}$ we find:
\begin{equation}
\label{beq1}
(\pa_t+v\cdot \nabla_x)f=
\lam^{-1} \int dv_2 \int_{S_+} dn  \quad (v-v_2)\cdot n \quad [f (x,v') 
f(x,v'_2)-f (x,v)f(x,v_2)]\;.
\end{equation}
The parameter $\lambda$ represents, roughly, 
the typical length a particle can cover without undergoing any collision ({\it mean free path}). (In \eqref {beq}
 we just chose $\lambda=1$.)
 \medskip
 
 {\em Remark.} After having performed the limit
 $N \to \infty$ and $\e \to 0$, there is no way to distinguish between incoming and outgoing pair velocities,
 because no trace of the parameter $\e$ is left in \eqref {beq1} and $n$ plays a role of random variable. 
 However keeping memory of the way we derived the Boltzmann equation, we shall 
 conventionally maintain the name
 `incoming' for velocities satisfying the condition $(v-v_2 )\cdot n\ge 0$ and consequently the pair 
 $ v',v_2'$ is outgoing in \eqref{beq1}.
 \smallskip

Equation \eqref {beq1} (or equivalently \eqref {beq}--\eqref {bcoll}) is the Boltzmann
equation for hard spheres. Such an equation has a statistical nature and it is 
not equivalent to the Hamiltonian dynamics from which it has been derived. Indeed
the $H$ theorem shows that it is not reversible in time in contrast with the laws of mechanics. 
Note that this is not the case for the Vlasov equation, which inherits all the properties of the Hamiltonian systems.   

From the analysis on the order of magnitude of the quantities in the game, we deduced that the Boltzmann equation
works in special situations only. The condition $N\e^2 =O(1)$ means that we consider a rarefied gas,
with almost vanishing volume density.
After Boltzmann established the equation, Harold Grad \cite {Gr49,Gr58} postulated its validity 
in the limit $N \to \infty$, $\e \to 0$ with $N\e^2 \to$  cost.\;as discussed above (this is often called, indeed, Boltzmann-Grad limit).

There is no contradiction in the irreversibility and the trend to equilibrium
obtained after the limit, when they are strictly speaking false for mechanical systems.
However the arguments above are delicate and require a rigorous,
deeper analysis. If the Boltzmann equation is not a purely
phenomenological model derived by assumptions {\it ad hoc} and justified by
its practical relevance, but rather a {\em consequence} of a mechanical model, we
should derive it rigorously. In particular the propagation of chaos should be
not an hypothesis, but the statement of a theorem.

After the formulation of the mathematical problem by Grad, Cercignani \cite{Ce72}
obtained the evolution equation (hierarchy) for the marginals of a hard-sphere system
and this was the starting point to derive rigorously the Boltzmann equation,
as accomplished by Lanford in his famous paper \cite {La75}, even though only for a short time interval.

Lanford's theorem is probably the most relevant result as regards the mathematical foundations
of kinetic theory. In fact it dissipated the many previous doubts on the validity of the Boltzmann equation
(although some authors refuse a priori the problem of deriving the equation starting from 
mechanical systems \cite{TM}).

Unfortunately, the short time limitation is serious. Only for special systems, as the case of
a very rarefied gas expanding in the vacuum, we can obtain a global validity result \cite{IP86,IP89}. 
The possibility of deriving the Boltzmann equation globally in time, at least in cases when we 
have a good global existence of solutions, is still an open and challenging problem.

\bigskip
We conclude this section with some historical remark. 

Before Boltzmann, Maxwell proposed a kinetic equation
that is nothing else than the Boltzmann equation integrated against test functions \cite {Ma67,Ma95}. He considered also more general
potentials, in particular inverse power law potentials, essentially for the special properties of their cross-section.

After Lanford's result, the case of smooth short-range potentials has been studied by other authors
\cite{Ki75,GSRT12,PSS13}, but the validity (or non-validity) of the Boltzmann equation in case of genuine long-range potentials is 
open.

A rigorous derivation of the hierarchy of equations for hard spheres as formally established by Cercignani is obtained
in \cite{Sp91,CIP94,Si13}.

\section*{  4. The Weak-Coupling limit and the Landau equation }

The Boltzmann equation is suited to the description of rarefied gases and one can ask
whether a useful kinetic picture can be applied also to the case of a dense gas. 
To introduce the problem, let us revisit first the Boltzmann-Grad limit in an alternative
way. Again we denote by $\e$ a small scale parameter denoting the ratio between the microscopic and
macroscopic scale. For instance the inverse number of atomic diameters necessary to cover $1$ meter, or
the inverse number of atomic characteristic times necessary to cover $1$ second. 
We scale by $\e$ space and time in the equations of motion (in the previous section, the hard-sphere hierarchy).
We need to specify the number of particles $N$. In a box of side one, 
there should be $N \approx \e^{-3}$ particles if one assumes that the intermolecular distance is of the same order of the molecular diameter. 
The number of collisions of a given particle per macroscopic unit time would be $\e^{-1}$.  
As we have seen, in a low-density regime $N$ scales differently, namely $N\approx \e^{-2}$,
the number of collisions per unit time is finite and the one-particle distribution function satisfies the Boltzmann equation.  

A variety of possible scalings describes different physical situations.
For instance the gas may be dense, $N=O(\e^{-3})$, and the particles {\em weakly interacting} via a {\em smooth} two-body potential $\varf$.  
To express the weakness of the interaction, we assume
that $\varf$ is rescaled by $\sqrt {\e}$. 
Since $\varf$ varies on a scale $\e$ (in macroscopic unities), the force
will be $O(\frac 1{\sqrt {\e}})$ and act on a time interval $O(\e)$.
The variation of momentum due to the single scattering is $O(\sqrt {\e})$ and the
number of particles met by a typical particle is $O(\frac 1{\e})$. Hence the
total momentum variation for unit time is $O(\frac 1{\sqrt {\e}})$.
However, in case of a homogeneous gas and symmetric forces, this variation should be
zero in the average. The computation of the variance leads to a result $\frac
1{\e} O(\sqrt {\e})^2=O(1)$. Therefore, based on a central limit type of argument, we expect 
that in the kinetic limit a diffusion equation in the velocity variable holds.
Moreover, even though the force induced by 
a given particle on a test particle is $O(\frac 1{\sqrt {\e}})$ (i.e.\;not small as in the mean-field limit), the fact 
that the total average variation of the momentum is small should be sufficient to ensure propagation of chaos.  

At the level of the kinetic equation (that is, assuming propagation of chaos), consider a model of collision
with operator of the form 
\begin{equation}
\label{colle}
Q(f,f)=\int dv_1 \int dp  \; w( p ) 
\delta
(p^2+(v-v_1)\cdot p) [f'f_1'-ff_1]
\end{equation}
where
$$
f'=f(v+p), \qquad f_1'=f(v_1-p).
$$
Here  $p$ is the transferred momentum in the collision and $w$ is the probability density
to have $p$ as an effect of the collision. The $\delta$ ensures energy conservation.

To express the fact that the
transferred momentum is small, let us rescale $w$ as $\frac 1{\var^3}w(\frac
p {\var})$ (so that the transferred momenta are $O(\e)$). 
In addition, let us rescale the inverse mean-free path by a
factor $\frac 1\var$ to take into account the high density of particles. The
collision operator becomes:
\begin{align}
Q_{\var}(f,f)& =  \frac 1{\var^4}\int dv_1 \int dp  \; w\left(\frac p {\var}\right) 
\delta
(p^2+(v-v_1)\cdot p) [f'f_1'-ff_1]\\
& = \frac 1{2\pi\var^2}\int dv_1 \int dp  \quad w( p
)\int_{-\infty}^{+\infty} ds\; e^{is (p^2 \var +(v-v_1)\cdot p)}\nn\\
&\ \ \ \ \times[f(v+\var p) f(v_1-\var p)-f(v) f(v_1)]\nn\\
&= \frac 1{2\pi\var}\int dv_1 \int dp  \quad w( p) \int _0^1 d\lam
\int_{-\infty}^{+\infty}ds \;e^{is (p^2 \var +(v-v_1)\cdot p)}\nn\\
&\ \ \ \ \times p\cdot (\nabla_v-\nabla_{v_1})f(v+\var \lam p) f(v_1-\var \lam p)\;.\nn
\end{align}
Here the smooth function $w$, which modulates the collision, is assumed 
to depend on $p$ through its modulus only. Note that we used a change of variables
$p/\e \to p$ and  the representation  formula in $\R^1$
$$
\d (x) = \frac 1 {2 \pi \e} \int_{-\infty} ^ {+\infty} ds \,e^{i \frac {sx}{\e}}.
$$

To outline the behaviour of $Q_{\var}(f,f)$ in the limit $\var \to 0$, we
introduce a test function $u$ for which, after a change of
variables (here $ (\cdot, \cdot)$ denotes the scalar product in
$L^2(dv)$):
\begin{align}
(u, Q_{\var}(f,f))&=\frac 1{2\pi\var}\int dv \int dv_1 \int dp  
\quad w( p) \int _0^1 d\lam
\int_{-\infty}^{+\infty}ds \;e^{is (p^2 (\var-2\var \lam) +(v-v_1)\cdot
p)}\\ &  \ \ \ \ \times u (v-\var \lam p) \quad p\cdot
(\nabla_v-\nabla_{v_1})ff_1=\nn \\
&=\frac 1{2\pi\var}\int dv \int dv_1 \int dp  
\quad w( p) \int _0^1 d\lam
\int_{-\infty}^{+\infty}ds\; e^{is (v-v_1)\cdot p}\nn\\
&\ \ \ \ \times
[u (v)-\var \lam p\cdot \nabla_v u (v) ]\quad
p\cdot(\nabla_v-\nabla_{v_1})ff_1\nn \\
&\ \ \ \ +\frac 1{2\pi}\int dv \int dv_1 \int dp  
\quad w( p) 
\int_{-\infty}^{+\infty}ds\; e^{is (v-v_1)\cdot p} u (v)\nn\\
&\ \ \ \ \times is\,p^2
\int _0^1 d\lam (1-2\lam) \quad p\cdot 
(\nabla_v-\nabla_{v_1})ff_1+ O(\var)\;.\nn
\end{align}

Note now that the term $O(\var ^{-1})$ vanishes because of the symmetry
$p\to-p$ ($w$ is even). Also the last term  is
vanishing, being null the integral in $d\lam$. As a result:
\begin{align}
(u,Q_{\var}(f,f))=&-\frac 1{4\pi}\int dv \int dv_1 \int dp  \; w(
p)
\int_{-\infty}^{+\infty}ds \; e^{is (v-v_1)\cdot p }\\
& p\cdot \nabla_v u \; p\cdot
(\nabla_v-\nabla_{v_1})ff_1+O(\var).\nn
\end{align}
Therefore we have formally recovered the kinetic equation 
\begin{equation}
\label{landau}
(\pa_t+v\cdot \nabla_x)f=Q_L (f,f)
\end{equation}
with a new collision
operator
\begin{equation}
Q_L(f,f)=\int dv_1 \nabla_v \; a (\nabla_v-\nabla_{v_1})ff_1,
\end{equation}
where $a=a(v-v_1)$ denotes the matrix
\begin{equation}
a_{i,j}(V)=\frac 12 \int dp \; w(p) \; \delta (V \cdot p) \;  p_ip_j. 
\end{equation}
This matrix can be handled in a better way by introducing polar
coordinates:
\begin{align}
a_{i,j}(V)&=\frac 1{2|V |}\int dp \; |p| w(p) \; \delta (\hat V \cdot
\hat p)
\; \hat p_i \hat p_j\\
& =\frac B{|V |}\int d \hat p  \; \delta (\hat V \cdot
\hat p)\; \hat p_i \hat p_j,\nn
\end{align}
where $\hat V$ and $\hat p$ are the versor of $V$ and $p$ respectively
and
\begin{equation}
B=\frac 12 \int_0^{+\infty} dr\; r^3\; w(r).
\end{equation}
Note that $B$ is the only parameter describing the interaction appearing
in the equation. Finally a straightforward computation yields:
\begin{equation}
a_{i,j}(V)=\frac B{|V |} \left(\delta_{i,j}-\hat V_i \hat V_j\right)\;.
\end{equation}

The collision operator $Q_L$ has been introduced by Landau in 1936 (\cite{La36,LL})
for the study of a weakly interacting dense plasma and \eqref {landau} is
called {\em Landau equation} (sometimes, Landau-Fokker-Planck).


The
qualitative properties of the solutions to the Landau equation are the
same as for the Boltzmann equation as regards the basic conservation
laws and the $H$ theorem.


Following the paradigm of the kinetic theory, we would like
to derive the Landau equation from particle systems.
A rigorous proof is however missing, even for short time intervals. We refer to \cite {BPS13}
for a partial result.

\section*{5. Some historical remarks}

The first attempt to implement the program of the kinetic theory is due to Boltzmann, who derived his celebrated equation for rarefied gases in 1872 \cite{Bo64}. He followed some previous work of Maxwell, who wrote a system of equations in 1867 for the moments of the velocity distribution in order to justify the equilibrium measure which inherits his name \cite{Ma67,Ma95}. 

Boltzmann's work was attacked by physicists and mathematicians due to the apparent basic contradiction between the $H$ theorem and the reversible nature of the Newton equations. In particular, the Poincar\'{e} recurrence theorem seemed to be in contrast with the approach to equilibrium.  
Boltzmann replied to the criticisms asserting that the equation has a statistical meaning. Of course he did not have at disposal the mathematical tools suitable to 
make this statement more precise.
We do not discuss further this interesting aspect \cite{C98}.

In spite of the success of the Boltzmann equation in solving practical problems concerning rarefied gases, the issue of a rigorous justification of the equation remained open for a long time.

As mentioned in Section 3, a significant step forward was done by Grad \cite{Gr49,Gr58}, who figured out the scaling limit in which the equation is expected to hold, in the framework of classical mechanics.
A second important contribution along Grad's approach was then obtained by Cercignani in 1972, when he established the hierarchy for the hard-sphere system whose first equation has been written in Section 3 \cite{Ce72}.  His analysis was formal but it opened the way to Lanford's rigorous result on the short time validity of the Boltzmann equation \cite{La75}.

Lanford's result solved the problem of conciliating the Boltzmann equation with the laws of classical mechanics. 

On the other hand, even in recent times, the Boltzmann equation has been often considered as a useful and successful tool of investigation, and not necessarily 
as a direct consequence of the principles of mechanics. This is, for instance, the position of Truesdell and Muncaster in their famous monograph \cite{TM}:

\medskip

{\it We will make no attempt to trace the source of this irreversibility in more general theories or physico-philosophical speculations. Rather, in the spirit of rational mechanics, we shall attempt to determine its specific and rigorous mathematical nature and consequences.}

\medskip

Another attitude was the one of the great probabilist Kac \cite{Kac,Kac2} who conceived a stochastic dynamics for an $N$-particle system
yielding rigorously, in the limit $N \to \infty$, the homogeneous Boltzmann equation. His work is contemporary to the one of Grad,
however the point of view is very different. A footnote in \cite{Kac} reads:


\medskip

{\it This formulation led to the well-known paradoxes which were fully discussed in the classical article of P. and T. Ehrenfest. These writers made it clear

\noindent (a) that the ``Stosszahlansatz" cannot be strictly derivable from purely dynamic considerations and

\noindent (b) that the ``Stosszahlansatz" has to be interpreted probabilistically. 

\noindent The recent attempts of Born and Green, Kirkwood and Bogoliubov to derive Boltzmann's equation from Liouville's equation and hence to justify the ``Stosszahlansatz" dynamically are, in our opinion, incomplete, inasmuch as they do not make it clear at what point statistical assumptions are introduced. 
 
 The ``master equation" approach which we have chosen seems to us to follow closely the intentions of Boltzmann. }

\medskip

The works quoted after point (b) were the first attempt of a justification of kinetic equations based on hierarchical techniques and the ``Stosszahlansatz" is the property of propagation of chaos necessary to close the hierarchy (we refer in particular to \cite{Bog62} for
a pioneering analysis including the three classical kinetic equations).



Therefore Kac's purpose is not just to provide a toy model, as it intends to be strongly related to the physics.
A further quotation form the same paper is:

\medskip

{\it Since the master equation is truly descriptive of the physical situation, and since existence and uniqueness of the solution of the master equation are almost trivial, the preoccupation with existence and uniqueness theorems for the  Boltzmann equation appears to be unjustified on grounds of physical interest and importance. }

\medskip

An important point is that Kac's model is restricted to homogeneous situations 
(no dependence on positions).
Interestingly enough, in the context of numerical simulations of rarefied gases, Bird constructed the successful scheme \cite{Bird} known as DSMC (direct simulation Monte Carlo), which splits the dynamics of a particle system into two parts: free motion + a stochastic interaction closely related to the one of Kac. In other words (without knowing Kac's work) Bird was providing an inhomogeneous stochastic model approximating the Boltzmann equation (see also \cite {CIP94} and references quoted therein).

Kac was greatly influenced by the famous treatise of Paul and Tatjana Ehrenfest \cite {EE15}, where the conceptual basis of statistical mechanics are discussed. Here the authors try to explain the nature of the Boltzmann equation and the emergence of irreversibility with the aid of simple examples.

We note, incidentally, that in \cite {EE15} a model is introduced (often called wind-tree model) in which a light (point) particle collides with a random distribution of square obstacles in the whole plane, in such a way that only four velocities are possible. The set of velocities is ${\cal V} =\{\pm e_1, \pm e_2\}$, where $e_i$, $i=1,2$ are the versors of the coordinate axes in the plane.   
An elastic collision of the light particle with an obstacle with sides oriented at $\pi/4$, is a rotation of the incoming velocity by $\pm \pi/2$. The corresponding kinetic equation is linear and has the form
$$
(\pa_t + v \cdot \nabla_x ) f =\frac 12 ( f(v^\perp)+f( -v^\perp)) -  f(v)
$$
where ${\cal V}\ni v  \to v^\perp$ is the rotation of $\pi /2$.

For the more realistic ``Lorentz model'' with circular obstacles and velocity set $S^1$, a rigorous derivation of the linear Boltzmann equation was 
obtained by Gallavotti in a remarkable paper \cite{G69}.  His approach applies as well to the Ehrenfest wind-tree model. 

We also mention a nonlinear version of the wind-tree model, namely the Broadwell model with kinetic equation
$$
(\pa_t + v \cdot \nabla_x ) f =( f(v^\perp)f( -v^\perp) -  f(v)f(-v) )\;.
$$
Surprisingly, this equation cannot be derived from the mechanical system of colliding square particles in the plane in the Boltzmann-Grad limit \cite {Ucy,CIP94}.
This counterexample shows how delicate can be a rigorous study of the low-density limit of deterministic systems.


\bigskip

Concerning dense gases and plasmas,
Landau proposed his kinetic equation in 1936.
He started from the Boltzmann 
equation with Coulomb cross-section, cutting divergences at short and at long distances. His argument is similar to the one presented here in Section 4
and the problem of the propagation of chaos is pragmatically avoided.

Bogolyubov works instead with the BBGKY and asserts that it would be necessary to obtain the Landau equation starting from particle systems under a suitable scaling limit instead of starting from the Boltzmann equation directly \cite{Bog62}. His discussion amounts to what is called nowadays the ``weak-coupling limit'' for the Landau equation (see also \cite{Bal}) and includes an attempt to outline the various regimes in which the kinetic equations are expected to be valid, starting again from the hierarchy.

\bigskip

The Vlasov equation was introduced first in 1938 \cite{Vla}, to study 
the time evolution of the distribution function of plasmas consisting of charged particles and long-range forces (for example, Coulomb),
in contrast with the Landau equation which is suited for particles interacting weakly via short-range forces. Actually both equations are needed to retain different aspects of the complicated dynamics of plasmas. 


\bigskip

We shall conclude by recalling the famous speech by Hilbert at the International Congress of Mathematicians in Paris 1900,
where he posed 23 problems as the basis of mathematical research in the forthcoming century \cite{H00}. 
Among these, the sixth is perhaps a less definite problem, but rather a broad field of investigation and a prophecy of the modern role of mathematics in physics.
It is titled {\em Mathematical Treatment of the Axioms of Physics} and reads:

\noindent 

\medskip
{\it The investigations on the foundation of geometry suggest $[\cdots]$
to treat in the same manner, by means of axioms, those physical sciences in
which mathematics plays an important part; in the first rank are the theory of
probabilities and mechanics}.

{\it As to the axioms of the theory of probabilities, it seems to me desirable that
their logical investigation should be accompanied by a rigorous and satisfactory
development of the method of mean values in mathematical physics, and in
particular in the kinetic theory of gases.

Important investigations by physicists on the foundations of mechanics are at
hand $[\cdots]$. Thus Boltzmann's work on the principles of mechanics suggests the
problem of developing mathematically the limiting processes, there merely
indicated, which lead from the atomistic view to the laws of the motion of
continua}.

\medskip

The necessity of a rigorous approach to the scaling limits starting from fundamental 
particle models is clearly expressed.
Moreover, the role of the mathematics in investigating
how different mathematical models of reality are connected is
outlined: 


\medskip

{\it $[\cdots]$ Further, the mathematician has the duty to test exactly in each instance whether
the new axioms are compatible with the previous ones. The physicist, as his
theories develop, often finds himself forced by the results of the experiments to
make new hypotheses, while he depends, with respect to the compatibility of the new
hypotheses with the old axioms, solely upon these experiments or upon certain
physical intuition, a practice which in the rigorously logical building up of a
theory is not admissible.} 

\medskip

Clarifying the scopes and methodologies of a physicist and a
mathematician establishes the role of the modern mathematical physics,
in which the concept of mathematical model, as a non-contradictory system
of axioms, is fundamental.


The 6th problem of Hilbert can nowadays be further specified, namely there are at least three kind of convergence
which can be analyzed: 1) Derive Boltzmann equation from particle systems;
2) Derive Euler and/or Navier-Stokes equations from the Boltzmann equation;
3) Derive Euler and/or Navier-Stokes equation from particle systems.


We have remarkable progress as regards points 1) and 2). The first point has been discussed in this note.
As regards point 2), Hilbert himself introduced an expansion which is the basic tool to derive the Euler equation for compressible fluids, in a suitable scaling limit, 
starting from the Boltzmann equation \cite{H16}. Many rigorous results deriving hydrodynamic laws from the Boltzmann equation have been obtained over recent years.
We underline that the hydrodynamic is the one of a perfect gas, since we start from a low-density regime. 
A more challenging problem is the derivation of the Euler equation from particle systems (point 3). The laws relating density, pressure and temperature are not those of a perfect gas, but they may be computed through the Gibbs state associated to the interacting potential of the system. Such a difficult problem is unsolved from the mathematical side. We mention only the formal computations of the pioneering work by Morrey \cite{Mo55}. Here the author gives a list of necessary steps to prove that the Euler equation can be obtained from the Newton laws. It is a notable work, having the merit to show in a logically clear way what is the link between the microscopic and macroscopic description of fluids. See \cite {EP04}  for a review on the argument.

\bigskip
\bigskip
\noindent {\bf Acknowledgments} We thank Thierry Paul for fruitful discussions.

\vskip1cm

\end{document}